\documentclass[fleqn,twoside,a4paper,dvips]{article}
\usepackage{espcrc2}
\usepackage{amsmath}
\usepackage{amssymb}
\usepackage{graphicx}
\usepackage[figuresright]{rotating}

\newcommand{\be}{\begin{equation}}
\newcommand{\ee}{\end{equation}}
\newcommand{\bea}{\begin{eqnarray}}
\newcommand{\eea}{\end{eqnarray}}
\newlength{\captionspace}
\title{Exploring the phase structure of lattice QCD with twisted mass
  quarks\thanks{Presented by C.~Urbach and F.~Farchioni.}}

\author{F.\ Farchioni\address{Institut f{\"u}r Theoretische Physik,
        Universit{\"a}t M{\"u}nster, Wilhelm-Klemm-Str.9, D-48149
        M{\"u}nster, Germany},
        C.\ Urbach\address[DESYZ]{NIC/DESY Zeuthen, Platanenallee 6, D-15738 Zeuthen,
        Germany}\address{Freie Universit{\"a}t Berlin,
        Institut f{\"u}r Theoretische Physik, Arnimallee 14, D-14196 Berlin, Germany},
        R.\ Frezzotti\address{INFN, Sezione di Milano and
        Dipartimento di Fisica, Universit{\`a} di Milano ``{\it Bicocca}'',
        Piazza della Scienza 3, I-20126 Milano, Italy},
        K.\ Jansen\addressmark[DESYZ],
        I.\ Montvay\address[DESYH]{Deutsches Elektronen-Synchrotron DESY, Notkestr.\,85,
        D-22603 Hamburg, Germany},
        G.C.\ Rossi\address{Dipartimento di Fisica, Universit{\`a} di  Roma
        ``{\it Tor Vergata}'' and INFN, Sezione di Roma 2, Via della Ricerca
        Scientifica, I-00133 Roma, Italy},
        E.E.\ Scholz\addressmark[DESYH],\\
        A.\ Shindler\addressmark[DESYZ],
        N.\ Ukita\addressmark[DESYH],
        I.\ Wetzorke\addressmark[DESYZ]
}

\begin{document}

\begin{abstract}
  The phase structure of zero temperature twisted mass lattice
  QCD is investigated.
  We find strong metastabilities in the plaquette observable when the
  untwisted quark mass sweeps across zero.
\vspace{1pc}
\end{abstract}

\maketitle

\section{Introduction}

In the process of approaching in lattice QCD the physical point, at
which the pion mass assumes its value as measured in experiment, the
simulation algorithms suffer from a substantial slowing down 
\cite{BERLIN,KARLTSUKUBA} which restricts present simulations to rather
high and unphysical values of the quark mass. Moreover the usual
Wilson-Dirac operator develops unphysical small eigenvalues at small
values of the quark mass which render the simulations more difficult
and sometimes even impossible.

\begin{figure}[t]
\centering
\includegraphics[width=.9\linewidth]{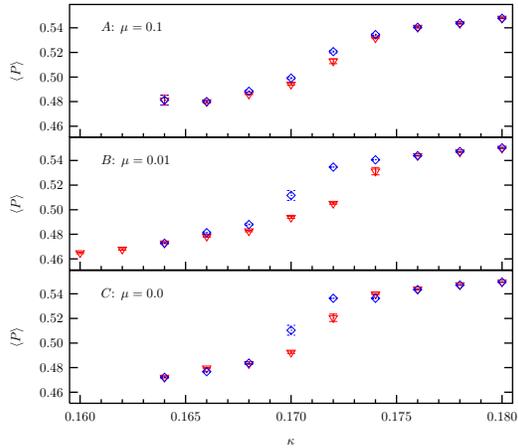}

\vspace{\captionspace}
\caption{
 Thermal cycles in $\kappa$ on $8^3\times 16$ lattices at $\beta=5.2$.
 The plaquette expectation value is shown for:
 $\mu=0.1$ (A); $\mu=0.01$ (B); $\mu=0$ (C).
 The triangles refer to increasing $\kappa$-values, the diamonds to
 decreasing ones. 
\label{cycles}}
\end{figure}

An elegant way out may be the use of Wilson twisted mass fermions
\cite{TMQCD} with the following fermionic action for $N_f = 2$ mass
degenerate flavors of quarks in the so called twisted basis ($\chi$)
\begin{equation}
  \label{eq:1}
  S[\chi,\bar\chi,U] = \bar\chi(D[U] + m_0 + \mu i \gamma_5\tau^3)\chi\, ,
\end{equation}
where $D[U]$ is the standard Wilson-Dirac operator, $m_0$ is the
untwisted quark mass parameter, $\mu$ is the twisted quark mass
parameter and $\tau^3$ is the third Pauli matrix acting in flavor space. In the present paper,
unless otherwise stated, the lattice spacing is set to unity: $a=1$.
The twisted mass $\mu$ serves as a natural infrared regulator for the
low lying eigenvalues of the Wilson twisted mass operator since 
\begin{equation}
  \label{eq:2}
  \begin{split}
    &\det(D[U] + m_0 + \mu i \gamma_5\tau^3) \\
    &= \det((D[U]+m_0)(D[U]+m_0)^\dagger+\mu^2)\, .
  \end{split}
\end{equation}
Note that on the l.h.s. the operator is the two flavor operator, while
on the r.h.s. $D[U]+m_0$ is only the one flavor part.
As for the gauge action, the usual Wilson plaquette action is used.
Note that the bare quark mass
$m_0$ is often represented by the hopping parameter $\kappa$ defined as
$\kappa = (2m_0+8)^{-1}$.

In addition to the infrared cut-off the Wilson twisted mass
 formulation allows to get full $O(a)$ improvement for
correlation functions and derived quantities with no need of
additional counterterms~\cite{FREZZOTTI-ROSSI,SCALING}, provided $m_0$ is set to its critical
value $m_{\rm crit}$ and the value of $\mu$ is kept constant as $a \to 0$.

Studying the phase structure of lattice QCD should be a pre-requisit
before starting to extract physical results. Indeed, in simulations of
$N_f = 3$ clover improved Wilson fermions \cite{JLQCD} and of $N_f = 2$
non-perturbatively improved Wilson fermions \cite{KARLTSUKUBA} metastabilities and
hysteresis effects were found. In this contribution, or in more detail
in \cite{Farchioni:2004us}, we discuss results for the phase structure of $N_f = 2$ Wilson
twisted mass fermions.

For our simulations of full QCD we have implemented two independent
algorithms: The Two-Step Multi-Boson algorithm (TSMB) and the Hybrid
Monte Carlo algorithm (HMC), both with even-odd preconditioning. For
the HMC we use in addition the Hasenbusch trick \cite{HASENBUSCH}. We
checked that we get the same results with both algorithms. 

We have performed simulations primarily at $\beta=5.2$, but we also have some data at
$\beta=5.3$ and $\beta=5.4$. The lattice sizes are $8^3\times16$,
$12^3\times24$ and $16^3\times32$.

\begin{figure}[ht]
\centering
\includegraphics[width=.9\linewidth]{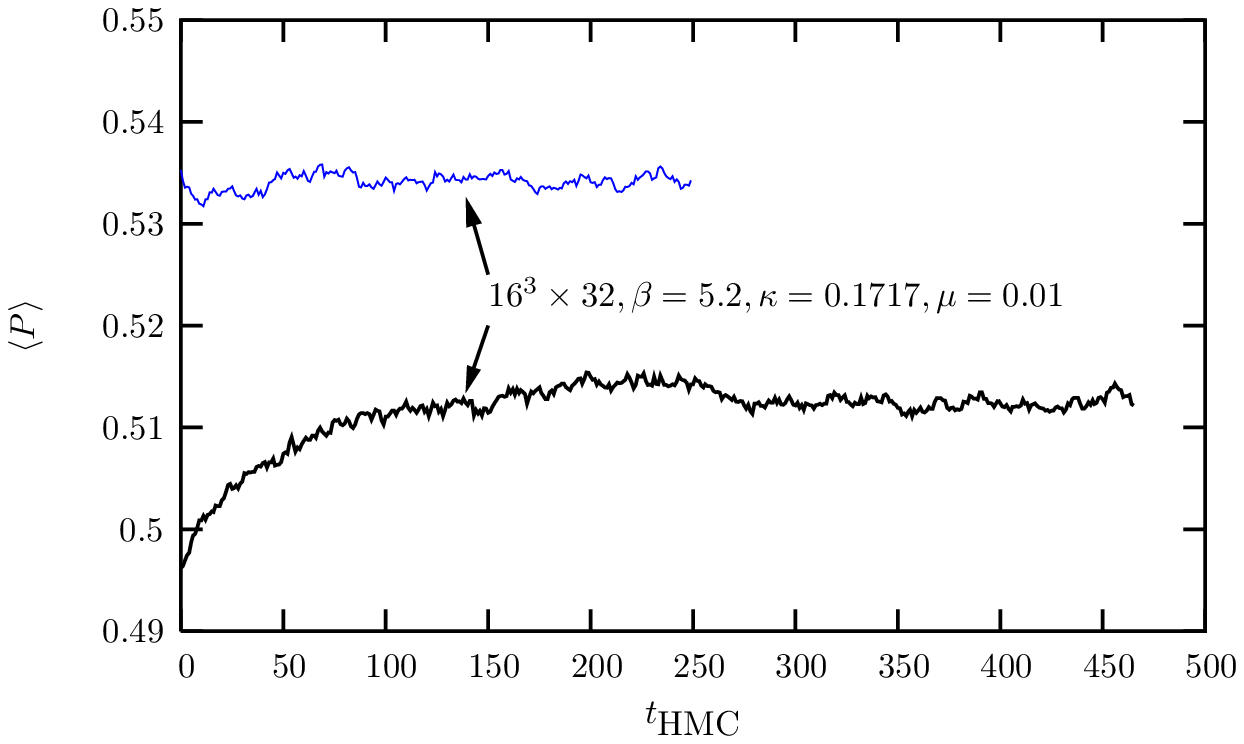}
\includegraphics[width=.9\linewidth]{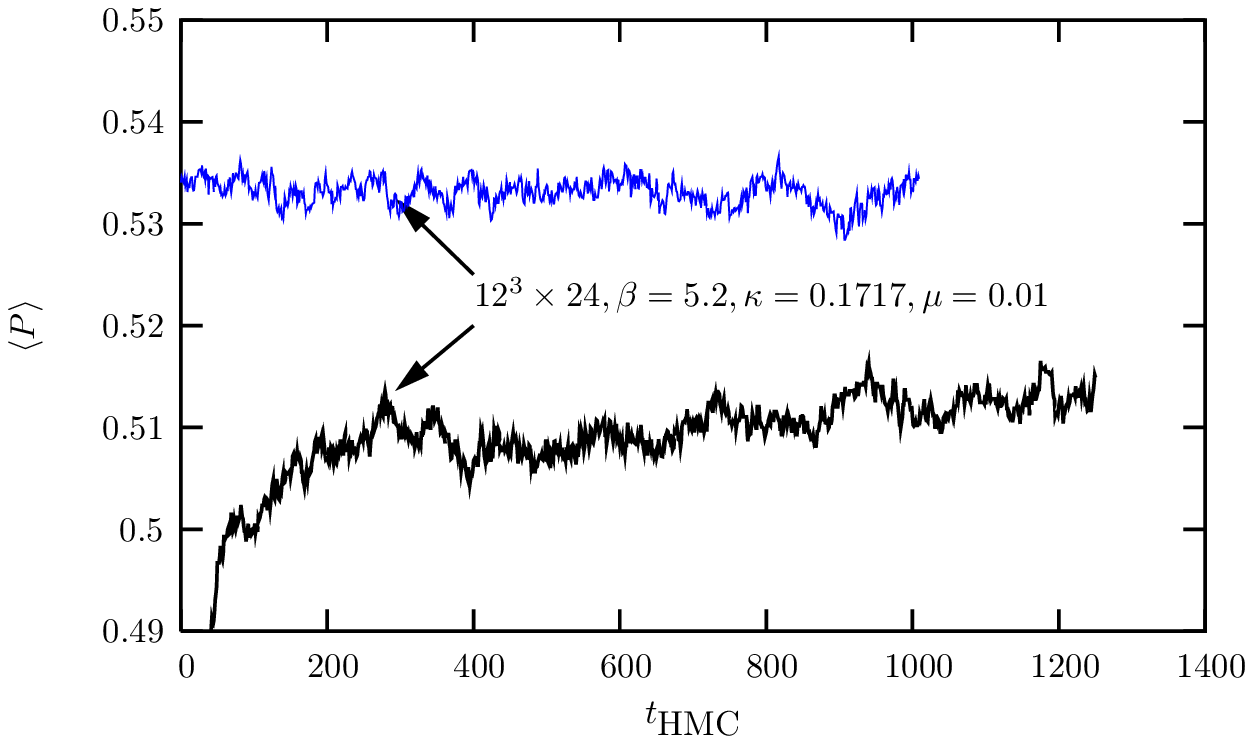}
\vspace{\captionspace}
\caption{
 Monte Carlo history of metastable states at $\beta=5.2$, $\mu=0.01$ and $\kappa=0.1717$. In the upper
 plot the lattice size is $16^3\times32$ and in the
 lower plot it is $12^3\times24$.
}
\label{metastabilities}
\end{figure}

\section{Thermal cycles and metastable states}

We started our investigation of the phase diagram 
by performing thermal cycles in $\kappa$. Fixing the values of
$\beta = 5.2$ and $\mu$ we incremented $\kappa$ from  a starting value
$\kappa_\textrm{start}$ until the final value $\kappa_\textrm{final}$ was reached
and then reversed the procedure. 
At each value of $\kappa$ $150$ configurations were produced -- without
performing further intermediate thermalization sweeps -- and
averaged over.

In fig.~\ref{cycles} we show three such thermal cycles, performed at
$\mu=0$, $\mu=0.01$ and $\mu=0.1$ from bottom to top.
In the cycles signs of hysteresis effects can be seen for $\mu=0$ and
$\mu=0.01$ while for the largest value of $\mu=0.1$ such effects are
hardly visible.
Hysteresis effects in thermal cycles {\em may be} signs of the
existence of a first order phase transition.
However, they should only be taken as first indications.

Guided by the results from the thermal cycles, we next performed
simulations at fixed values of $\mu$ and $\kappa$, starting with
ordered and disordered configurations, staying again at $\beta=5.2$.

In fig.~\ref{metastabilities} we show the Monte Carlo time evolution
of the plaquette expectation value for two sets of parameters at $\beta=5.2$. In the
upper plot the lattice size is $16^3\times32$ and in the lower plot the
lattice size $12^3\times24$, both with $\mu=0.01$ and $\kappa=0.1717$.
We find coexisting branches with different values of the average
plaquette with a rather large gap in between. As can be seen the gap
is not decreasing with increasing lattice size and therefore 
this behavior cannot be ascribed to any finite volume effect.
Furthermore we observed this phenomenon in simulations with $\mu=0$
and with both algorithms. We therefore conclude that the 
existence of metastable states is a generic feature of lattice QCD in this
formulation.

\section{Pseudoscalar and quark masses}

\begin{figure}
\centering
\includegraphics[width=.9\linewidth]{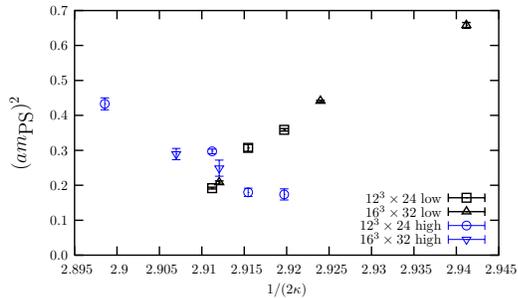}
\vspace{\captionspace}
\caption{
  Pseudoscalar mass as a function of $1/(2\kappa)$ at $\beta=5.2$ and $\mu=0.01$.
}
\label{pionmass}
\end{figure}

In order to study the physical properties in the two metastable states
we measured the (charged) pseudoscalar meson mass and the untwisted PCAC quark
mass. We obtained the pseudoscalar mass from the pseudoscalar
correlation function in the $\chi$-basis while we measured the untwisted
PCAC quark mass from the axialvector current in the $\chi$-basis: 
\begin{equation}\label{mpcac}
m_\chi^{\mathrm{PCAC}} \equiv
\frac{\langle\partial_\mu^*\bar\chi\gamma_\mu\gamma_5
\frac{\tau^\pm}{2}\chi(x)\;  \hat O^\mp(y)\rangle}
{2\langle\bar\chi\frac{\tau^\pm}{2}\gamma_5\chi(x)\; 
\hat O^\mp(y)\rangle} \ .
\end{equation}
Here $\hat O^\mp$ is a suitable operator that we have chosen to be
the pseudoscalar density
$\hat O^\mp = \bar\chi\frac{\tau^\mp}{2}\gamma_5\chi(x)$,
$\partial_\mu^*$ is the lattice backward derivative defined as usual
and $\tau^\pm = \tau_1\pm i\tau_2$. 
One can show that in the continuum limit $a\to 0$ the quantity
$m_\chi^{PCAC}$ is asymptotically proportional to  $m_0-m_{\textrm{crit}}$.

\begin{figure}[t]
\centering
\includegraphics[width=.9\linewidth]{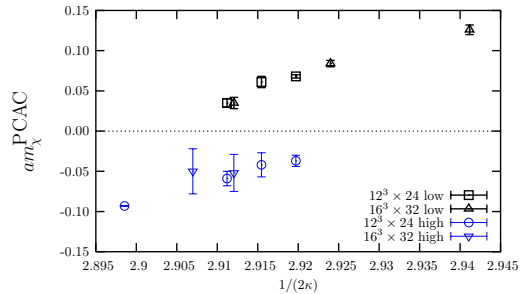}
\vspace{\captionspace}
\caption{
  $m_\chi^{\textrm{PCAC}}$ as a function of $1/(2\kappa)$ at $\beta=5.2$ and $\mu=0.01$.
}
\label{quarkmass}
\end{figure}

In fig.~\ref{pionmass} we show the pseudoscalar mass in lattice units as
function of $1/(2\kappa)$.
We observe that the ``pion'' mass is rather large and the most striking
effect in the graph is that it can have two different values at the
same $\kappa$.
If we consider the quark mass $m_\chi^{\mathrm{PCAC}}$ in
fig.~\ref{quarkmass}, we see that in the phase with a low plaquette expectation value
the mass is positive while for high values of the plaquette
expectation it is negative. Values of $m_\chi^\textrm{PCAC}$ with
opposite sign coexist for some values of $\kappa$. From this fact one
can safely argue that the value of $1/(2\kappa_\textrm{crit})$ lies
between $2.91$ and $2.92$.

Figures~\ref{metastabilities}-\ref{quarkmass} clearly reveal that for $\mu=0.01$
metastabilities show up in the quantities we have
investigated, such as $m_\textrm{PS}$, $m_\chi^{PCAC}$ and the average
plaquette, if $m_0$ is close to its critical value.

\section{Determination of the twist angle}

The knowledge of the twist angle $\omega$, the polar angle in the
plane of the untwisted and twisted mass, is important e.g. when comparing lattice data
with analytical predictions~\cite{TMChPT} from Wilson chiral perturbation
theory~(WChPT). 
We present here a method which allows to determine the
twist angle only on the basis of symmetry.

Following~\cite{TMQCD}, we introduce\footnote{In~\cite{TMQCD} this
definition of the twist angle was called  $\alpha$.} the twist angle $\omega$ as the chiral
rotation angle between the renormalized (physical) chiral
currents $\hat{V}^{a}_\mu$, $\hat{A}^{a}_\mu$  and the bare bilinears of 
the $\chi$-fields $V^{a}_\mu$, $A^{a}_\mu$ with
renormalization constants $Z_V$ and $Z_A$. Thus we have
\bea\label{v1}
\hat{V}^{a}_\mu=Z_V V^{a}_\mu\, \cos\omega\,  +\epsilon_{ab} \, Z_A A^{b}_\mu\, \sin{\omega}\,\\\label{v2}
\hat{A}^{a}_\mu=Z_A A^{a}_\mu\, \cos\omega\,  +\epsilon_{ab} \, Z_V V^{b}_\mu\, \sin{\omega}\, ,
\eea
where only charged currents are considered ($a$=1,2). 
The twist angle $\omega$ is related to the ratio
of the renormalized twisted and untwisted masses
entering the chiral Ward identities~\cite{TMQCD}:
\be\label{relomega}
\omega=\arctan{(\mu_R/m_R)}\ .
\ee
We define in addition the two auxiliary angles
\bea\label{o1}
\omega_V=\arctan (Z_AZ_V^{-1}\tan\omega)\ ,\\\label{o2}
\omega_A=\arctan (Z_VZ_A^{-1}\tan\omega)\ .  
\eea
In terms of $\omega_V$, $\omega_A$ eqs.~(\ref{v1}), (\ref{v2}) are written
\bea
\hat{V}^a_\mu=\!\ {\cal N}_V\, (\cos\omega_V V^a_\mu +\epsilon_{ab}\sin{\omega}
_{V} A^b_\mu)\ \label{renv}\\\label{rena}
\hat{A}^a_\mu=\!\ {\cal N}_A\, (\cos\omega_A A^a_\mu +\epsilon_{ab}\sin{\omega}
_{A} V^b_\mu)\ ,
\eea
where the overall multiplicative renormalization reads
\be\label{norm}
{\cal N}_{X}=\frac{Z_X}{\cos\omega_X \sqrt{1+\tan{\omega}_V\tan{\omega}_A}}\ .
\ee
From (\ref{o1}), (\ref{o2}) it follows:
\be\label{missing} 
\omega=\arctan \left(\sqrt{\tan{\omega}_V\tan{\omega}_A}\right)\ .
\ee
$\omega_V$ and $\omega_A$ can be directly determined by imposing parity conservation
for suitable matrix elements; e.g., with 
$P^{\pm}(x)=\bar\chi\frac{\tau^\pm}{2}\gamma_5\chi(x)$:
\be
\sum_{\vec{x},\vec{y}} \langle \hat{A}^+_0(x)\hat{V}^-_0(y)\rangle\! =\! 
\sum_{\vec{x},\vec{y}} \langle \hat{V}^+_0(x)P^-(y)\rangle = 0\ .
\ee
These equations admit the solution
\bea
\tan{\omega}_A=\ \ \ \ \ \ \ \ \ \ \ \ \ \ \ \ \ \ \ \ \ \ \ \ \ \ \ \ \ \ \ \
\ \ \ \ \ \ \ \ \ \ \ \ \ \ \ \  
\nonumber\\
\frac
{
  i\!\sum_{\vec{x},\vec{y}}\langle {A^+_0}(x) V^-_0(y)\rangle
  \!+\!\tan{\omega}_V\!
    \sum_{\vec{x},\vec{y}}\langle {A^+_0}(x) A^-_0(y)\rangle
  } 
  {
   \sum_{\vec{x},\vec{y}}\langle {V^+_0}(x) V^-_0(y)\rangle
  \!-\!i\!\tan{\omega}_V\!
    \sum_{\vec{x},\vec{y}}\langle {V^+_0}(x) A^-_0(y)\rangle
  } \nonumber
\eea
\be
\tan{\omega}_V=
\frac{-i\sum_{\vec{x},\vec{y}}\langle V^+_0(x) {P^-}(y) 
  \rangle} {\sum_{\vec{x},\vec{y}}\langle  A^+_0(x) {P^-}(y) \rangle}
\ . \ \ \ \ \ \ \ \ \ \ \ 
\ee
Alternatively, a determination of $\omega_V$ and $\omega_A$ is
given by the vector and axialvector Ward identities, respectively;
e.g. in the vector case, by enforcing the Ward identity with the
insertion of some appropriate operator $\hat{O}(x)$: 
$\langle \partial_\mu^*\hat{V}_\mu^+(x){\hat{O}}^-(y)\rangle$=0.

Once $\omega_V$ and $\omega_A$ are determined, the twist angle $\omega$ is
obtained by eq.~(\ref{missing}). 
The method described above for determining the twist angle can also be used
in case of simulations with partially quenched twisted
mass quarks. The estimate of $\omega$ is of course affected by $O(a)$
ambiguities.
For $\mu\!\propto\!\mu_R\!\neq\!0$, $|\omega|\!=\!\pi/2$ signals
$m_0\!=\!m_\textrm{crit}$. 

\begin{figure}[t]
\includegraphics[width=.9\linewidth,angle=-90]{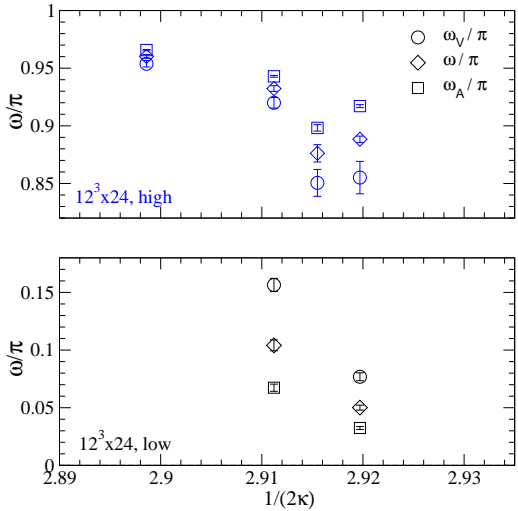}
\vspace{\captionspace}
\caption{
  Twist angle as a function of $1/(2\kappa)$ at $\beta=5.2$ and $\mu=0.01$ in
  the two branches.
}
\label{omega}
\end{figure}

We determined the twist angle for sets of configurations at $\beta$=5.2, $\mu$=0.01 in 
the positive and negative quark mass branches; 
results are reported in fig.~\ref{omega}. 

Owing to eq.~(\ref{relomega}) $\omega$  should approach the value $\omega=\pi/2$ 
from above or below, respectively, when the untwisted quark mass $m_\chi^{PCAC}\!\propto\!m_R$
gets close to zero from negative or positive values and $\mu\!\propto\!\mu_R$ is
kept fixed to a nonzero value. For $\mu =0.01$, outside the region of metastabilities, we observe 
 a trend of $\omega$ consistent with the above
expectation (see fig.~\ref{omega}). In the metastability region $(2\kappa)^{-1} =2.91-2.92$ the measured 
values of $\omega$ lie above or below $\pi/2$ in the high or low plaquette phase,
respectively, reflecting the behavior of $m_\chi^{PCAC}$ in fig.~\ref{quarkmass}.
The values of $\omega$ are far away from $\pi/2$.

\begin{figure}[t]
\includegraphics[width=.8\linewidth,height=0.9\linewidth,angle=-90]{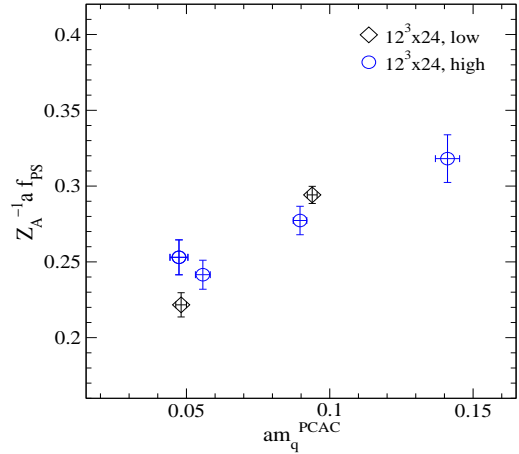}
\vspace{\captionspace}
\caption{
  Pseudoscalar decay constant as a function of the physical PCAC quark mass
  at $\beta=5.2$ and $\mu=0.01$ in the two branches.
}
\label{fpi_vs_mq}
\end{figure}

Eqs.~(\ref{renv})-(\ref{norm}) allow to determine 
the physical currents  $\hat{V}^a_\mu$ and $\hat{A}^a_\mu$ up to the
usual multiplicative renormalizations $Z_V$ and $Z_A$.
The physical PCAC quark mass was computed from the Ward identity for the
physical axialvector current, analogously to~(\ref{mpcac})
\begin{equation}\label{mpcac_full}
m_q^{\mathrm{PCAC}} \equiv
\frac{\langle\partial_\mu^*Z_A^{-1}\hat{A}_\mu^\pm(x)\;  \hat O^\mp(y)\rangle}
{2\langle P^{\pm}(x)\;\hat O^\mp(y)\rangle} \ .
\end{equation}
We determined the pion decay constant $f_\pi$ from the physical axialvector
current. The results are plotted in fig.~\ref{fpi_vs_mq} versus 
the physical PCAC quark mass.

\section{Interpretation of the results}
The observed metastabilities characterizing coexisting phases
with opposite sign of the untwisted quark mass support the picture of a first
order phase transition (PT) in the $m_0$-$\mu$ plane for small
values of $\mu$ and $m_0$ close to $m_\textrm{crit}$.
Also the observed gap in the average plaquette finds a natural explanation~\cite{Farchioni:2004us} in the 
presence of a first order PT, when the effects of the spontaneous breaking of
chirality are considered in combination with the explicit breaking of
the symmetry due to the Wilson term in the fermionic action.
 
A first order PT for Wilson fermions at small 
$O(a^2)$ quark masses is one of the two possible scenarios 
predicted by chiral perturbation theory with inclusion of lattice corrections~\cite{SHARPE-SINGLETON}
(WChPT). The other one is the well-known Aoki phase, known to be realized
at strong gauge couplings (see \cite{HUMBOLDT}).
Which of the two scenarios applies depends on the sign of the second
order ($O(a^2)$) corrections in the effective potential.
The generalization of the results of ref.~\cite{SHARPE-SINGLETON} to the
case of twisted mass fermions was recently worked out
independently in~\cite{Munst,SCORZATO} and~\cite{ShaWu}: the first order PT
extends to $\mu\neq0$ getting weaker and weaker for increasing
$\mu$; at the endpoint $\mu_c=O(a^2)$ the PT becomes a critical point.

The metastability of two phases with positive and negative untwisted quark mass has an 
interpretation, at a microscopic level, in terms of the properties of the eigenvalue spectrum 
of the Wilson-fermion matrix; studies in this direction are in
progress~\cite{PROGRESS} (see~\cite{QQ+Q1} for a numerical study of the case  $\mu$=0).
The qualitative picture~\cite{PROGRESS} is that the tunneling
between the two phases implies a massive rearrangement of small eigenvalues.
This process is suppressed by the zero of the fermion determinant when the
eigenvalues move close to the origin in the complex plane.
For $\mu$$\neq$0 and large enough, the metastability is lifted since 
the condition $|{\bf Im}\lambda|\geq\mu$ implies the depletion of the spectrum
around the origin. This conclusion is in agreement with the prediction of
twisted mass WChPT (see above).

\section{Conclusion and outlook}

We investigated the zero temperature phase diagram of twisted mass Wilson fermions.
The inclusion of a twisted mass term made it easier to explore
the region of small (untwisted) quark masses. Metastabilities, signaled by
different values of the plaquette, were visible in this region 
in thermal cycles and long-living metastable states.
The two metastable branches are characterized by opposite values of the
untwisted quark mass. The observed phase transition is rather strong: instabilities start already
for heavy ($\sim$740 MeV) pion masses. This is also reflected in the values of 
the twist angle in the two metastable branches, which are far from $\omega=\pi/2$.

We interpret our findings in terms of the existence of a first order
phase transition line in lattice QCD with Wilson fermions; this 
scenario is supported by Wilson chiral perturbation
theory. We argue that a phase transition for small quark masses is
accompanied at a microscopic level with the properties of the spectrum of the
Wilson-fermion matrix in proximity of the origin in the complex
plane.

The existence of a first order phase transiton line for small quark masses
poses the question, how a critical bare mass can be estimated in practice.
One possible choice is to identify $m_\textrm{crit}$ with the value of $m_0$ where the
pion mass is minimal. With the untwisted mass set to this value and for $\mu > \mu_c=O(a^2)$
the vacuum is unique.

There are still many aspects which must be clarified in order to get a more
detailed picture. First, how the present metastabilites are related
with those observed in the literature with different kind of actions and
number of flavors~\cite{BLUMETAL,JLQCD,KARLTSUKUBA} (see also~\cite{JLQCD2} for a recent
discussion). We shall start by investigating the influence of the
lattice gauge action on the phase structure near zero quark mass~\cite{PROGRESS}.
Preliminary results at lattice spacing $a \simeq 0.2\,{\rm fm}$ show
that replacing the Wilson plaquette action by the DBW2 action makes
the minimal pion mass and the jump of the average plaquette substantially
smaller.

Another direction of investigation is the dependence of the metastability
on the gauge coupling $\beta$. In the continuum limit
($\beta\to\infty,\;a\to 0$) the minimal pion mass and the jump of the
average plaquette are expected to vanish and the first order phase
transition line is expected to shrink to a first order phase transition point.

Finally, the most important question is how phenomenology can be done with
Wilson fermions in presence of the metastability. 
In other terms, one has to determine the lightest quark mass that
can be simulated in a stable phase, for values of lattice spacings accessible
to present simulations.

\vspace{-.25 cm}

\section{Acknowledgments}
F.F. and C.U. would like to thank the {\it Deutsche Forschungsgemeinschaft} for
funding their attendance at the conference.


\begin{thebibliography}{9}
%
\bibitem{BERLIN}
C.~Bernard {\it et al.},
Nucl.\ Phys.\ Proc.\ Suppl.\  {\bf 106}, 199 (2002).
%
\bibitem{KARLTSUKUBA}
K.~Jansen,
Nucl.\ Phys.\ Proc.\ Suppl.\  {\bf 129}, 3 (2004); hep-lat/0311039.
%
\bibitem{TMQCD}
R. Frezzotti {\it et al.},
Nucl. Phys. Proc. Suppl. {\bf 83} (2000) 941; hep-lat/9909003; \\
R. Frezzotti {\it et al.},
JHEP {\bf 0108} (2001) 058; hep-lat/0101001.
%
\bibitem{FREZZOTTI-ROSSI}
R.\ Frezzotti and G.C.\ Rossi, JHEP {\bf 0408} (2004) 007;
hep-lat/0306014; \\
Nucl.\ Phys.\ Proc.\ Suppl. {\bf 128} (2004) 193; hep-lat/0311008;
R.~Frezzotti, talk at this conference.
%
\bibitem{SCALING}
K.~Jansen {\it et al.} [XLF Collaboration],
Phys.\ Lett.\ B {\bf 586}, 432 (2004); hep-lat/0312013; \\
A.~Shindler, talk at this conference.
\bibitem{JLQCD}
JLQCD Collaboration, S. Aoki et al.,
Nucl. Phys. Proc. Suppl. {\bf 106} (2002) 263; hep-lat/0110088.
%
\bibitem{Farchioni:2004us}
F.~Farchioni {\it et al.}, hep-lat/0406039.
%
\bibitem{HASENBUSCH}
M. Hasenbusch,
Phys. Lett. {\bf B519} (2001) 177; hep-lat/0107019.
%
\bibitem{TMChPT}
G.~M{\"u}nster and C.~Schmidt, Europhys.\ Lett.\  {\bf 66} (2004) 652,
hep-lat/0311032; G.~M{\"u}nster, C.~Schmidt and E.~E.~Scholz, hep-lat/0402003.
%
\bibitem{Munst}
G.~M{\"u}nster, talk at this conference and hep-lat/0407006.
%
\bibitem{SCORZATO}
L. Scorzato, hep-lat/0407023.
\bibitem{SHARPE-SINGLETON}
S.R. Sharpe and R. Singleton, Jr.,
Phys. Rev. {\bf D58} (1998) 074501; hep-lat/9804028.
%
\bibitem{HUMBOLDT}
E.M. Ilgenfritz {\it et al.},
Phys. Rev. {\bf D69} (2004) 074511; hep-lat/0309057.
%
%
\bibitem{ShaWu}
S.~R.~Sharpe and J.~M.~S.~Wu, talk at this conference; hep-lat/0407025; hep-lat/0407035.
%
\bibitem{PROGRESS} 
F.~Farchioni {\it et al.}, work in progress.
%
\bibitem{QQ+Q1}
qq+q Collaboration, F. Farchioni {\it et al.},
Eur. Phys. J. {\bf C26} (2002) 237; hep-lat/0206008.
\bibitem{BLUMETAL}
T. Blum et al.,
Phys. Rev. {\bf D50} (1994) 3377; hep-lat/9404006.
%
\bibitem{JLQCD2}
S.~Aoki {\it et al.}  [JLQCD Collaboration], hep-lat/0409016.
%
\end{thebibliography}
\end{document}